# Naturally occurring van der Waals materials

Riccardo Frisenda,[1] Yue Niu,[2] Patricia Gant,[1] Manuel Muñoz[3] and Andres Castellanos-Gomez*[1]

[1] Materials Science Factory, Instituto de Ciencia de Materiales de Madrid, Consejo Superior de Investigaciones Científicas, 28049, Madrid, Spain

[2] National Center for International Research on Green Optoelectronics and Guangdong Provincial Key Laboratory of Optical Information Materials and Technology, Institute of Electronic Paper Displays, South China Academy of Advanced Optoelectronics, South China Normal University, Guangzhou 510006, People's Republic of China.

[3] Instituto de Tecnologías Físicas y de la Información ITEFI-CSIC, 28006 Madrid, Spain

*E-mail: andres.castellanos@csic.es

**ABSTRACT:**.

The exfoliation of two naturally occurring van der Waals minerals, graphite and molybdenite, arouse an unprecedented level of interest by the scientific community and shaped a whole new field of research: 2D materials research. Several years later, the family of van der Waals materials that can be exfoliated to isolate 2D materials keeps growing, but most of them are synthetic. Interestingly, in nature plenty of naturally occurring van der Waals minerals can be found with a wide range of chemical compositions and crystal structures whose properties are mostly unexplored so far. This Perspective aims to provide an overview of different families of van der Waals minerals to stimulate their exploration in the 2D limit.

## Introduction

The mechanical exfoliation of graphite to isolate single-layers of graphene [1] has been closely followed by intense experimental efforts to exfoliate other layered bulk crystals to achieve a broad catalogue of 2D materials with complementary properties to those of graphene.[2–5] The exploration of other 2D materials is motivated not only by the need of finding materials whose properties are optimal for a certain specific application but also by the fact that along with this exploratory process one might often find interesting (and sometimes even unexpected) physical phenomena.

Interestingly, despite the fact that 2D materials research was triggered by the exfoliation of naturally occurring graphite minerals [1] and that these natural layered minerals proved to be a readily source of high quality 2D material with extremely high charge mobilities,[6] the amount of naturally occurring layered materials studied so far is still very scarce (mainly limited to molybdenite,[7,8] tungstenite,[9] muscovite[10–12] and clays[13,14]). On the other hand, the amount of works on exfoliation of synthetic layered materials has kept growing: BN, $MoSe_2$,$WSe_2$, BP, $TiS_3$, SnS, $SnS_2$, InSe, $In_2Se_3$, GaSe, GaTe, $ReS_2$, $ReSe_2$, $NbSe_2$, $TaS_2$, among others.[4,15–33]

The goal of this manuscript is to present an overview, rather than a comprehensive study, of different families of naturally occurring van der Waals materials (many of whom are almost unexplored so far) to provide a starting point for future works on these natural 2D materials and to stimulate further studies on other layered minerals. Moreover, from the study of naturally occurring layered minerals one might get inspired to develop new synthesis approaches to grow synthetic materials whose structure mimics that of a natural layered mineral family.

In the following we will introduce illustrative examples of natural van der Waals materials belonging to different mineral families: natural elements, sulfides, sulfosalts, oxides, silicates, phosphates and carbonates.

## Natural elements

Highly crystalline graphite mineral rocks are found in nature and are currently used by many research groups to fabricate single-layer graphene by mechanical or liquid phase exfoliation routes. For example, natural graphite source has been employed in some of the seminal graphene papers to unravel unexplored physical phenomena like a new type of quantum Hall effect or the fundamental relationship between the transparency of graphene and the fine structure constant.[34–36] Figure 1a shows a schematic of the crystal structure of graphite where the different layers, that are held together by weak van der Waals interaction, are visible. Each layer is formed by covalently bonded $sp^2$ hybridized carbon atoms arranged in a honeycomb lattice. Figure 1b shows a photograph of a natural graphite piece and 1c displays an optical microscopy image of a few-layer graphene flake extracted from the mineral shown in Figure 1b by mechanical exfoliation and transferred onto a 285 nm $SiO_2$/Si substrate by all-dry transfer.[37–39]

Other examples of natural elemental van der Waals materials, although much less abundant than graphite, are native bismuth, antimony, selenium and tellurium, see Figure 1d to 1k. While bismuth and antimony forms 2D layers (slightly puckered), selenium and tellurium constitute an interesting example of van der Waals material with a quasi 1D structure. In fact, these materials are formed by the van der Wals interaction between neighbouring helical wires of Se- ot Te- atoms forming a crystal. Regarding the electrical properties, bismuth is a metal (becomes a direct band gap semiconductor with 0.2-0.3 eV gap in the single- and bi-layer limit),[40] antimony a semi-metal (becomes an indirect band gap semiconductor with 2.28 eV in the single-layer limit)[41] and selenium and tellurium are semiconductors with band gaps of 0.31 eV (indirect) and 2 eV (direct).[42–44]Although there are recent works on atomically thin bismuth, antimony, selenium and tellurium (and the interest of the community in these systems is rapidly growing) to our knowledge all these works employ synthetic materials as the starting point for the exfoliation or synthesis of the nanolayer systems that they study.[42–51] For example, an hydrothermal synthesis method have been used to grow tellurium nanosheets with thickness down to 0.4 nm (one unit cell). These nanosheets were applied in field effect transistors that show a high p-type mobility of up to 600 $cm^2V^{-1}s^{-1}$ and a high anisotropic in-plane electrical transport (anisotropy ratio ~1.5) and photodetectors with a responsitivy of 8 A/W for 2.4 µm and a cut-off wavelength of 3.4 µm.[44,52] Selenium nanosheets as thin as 5 nm were also synthesized by vapour transport and they have been applied in phototransistors showing a p-type character with mobilities up to 0.26 $cm^2V^{-1}s^{-1}$ and photoresponsivities reaching 260 A/W.[42]

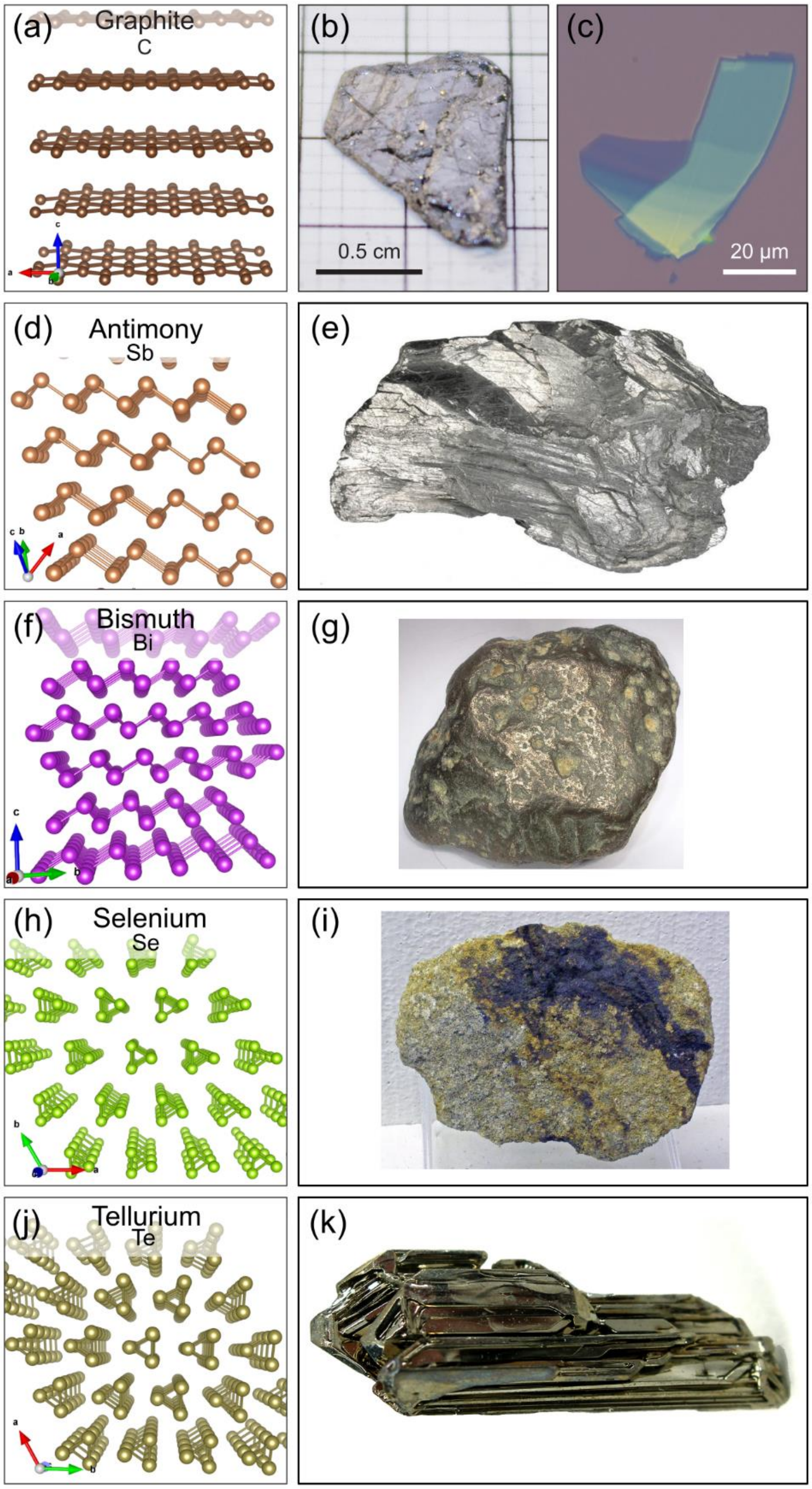

**Figure 1: Elemental van der Waals minerals.** (a), (d), (f), (h) and (j) 3D representation of the crystal structure of graphite,[147] native antimony,[148] native bismuth,[148] native selenium[149] and native tellurium.[150] (b), (e), (g), (i) and (k) Pictures of mineral rocks of graphite, native antimony,[151] native bismuth,[152] native selenium[153] and native tellurium,[154] respectively. (c) Optical microscopy image of a few-layer graphene flake mechanically exfoliated from the bulk graphite mineral shown in (b).

## Sulfides

Molybdenite mineral (with formula $MoS_2$) is very abundant in nature in its 2H polytype, in which the different layers are stacked in ABA fashion, and high-quality large crystals are easily available. Bulk molybdenite is an n-type indirect bandgap semiconductor with a band gap of ≈1.3 eV which becomes a direct semiconductor with 1.85-1.90 eV gap when thinned down to a single-layer.[53,54] Figure 2a, 2b and 2c show respectively the crystal structure of 2H-molybdenite, a picture of a piece of 2H-molybdenite mineral, and an optical microscopy image of a flake mechanically exfoliated from the mineral and onto a 285 nm $SiO_2$/Si substrate.

The combination of high carrier mobility (40-120 $cm^2V^{-1}s^{-1}$ at room temperature and up to $10^4$ $cm^2V^{-1}s^{-1}$ at 4K),[6,55] high photoresponse (up to ~1000 A/W) [56,57] and remarkable mechanical resilience of $MoS_2$ (breaking at strains of 6-11%) [58–61] have spurred the research on this material in the last 10 years. In fact, mechanically exfoliated flakes of natural molybdenite have been extensively used to fabricate several kinds of electronic and optoelectronic devices like field effect transistors,[8] photodetectors,[62–64] solar-cells,[65,66] non-volatile memories,[67,68] logic circuits among others.[3,67–69]

Although way less abundant than the 2H polytype, molybdenite can also be found in nature in its 3R polytype. In fact, 3R-molybdenite is typically found as a micromineral (small pieces inside a matrix of another rock). This mineral differs from the 2H-molybdenite in the way the layers are stack on top of each other, having a ABCA stacking (see a representation of its crystal structure in Figure 2d). This polytype is interesting because of its particular stacking of the layers that yields to a broken inversion symmetry even in bulk form (while in the case of the 2H polytype only an odd number of monolayers does not present inversion symmetry). [70,71] The lack of an inversion center symmetry is particular important for valleytronics because it introduces K-valley dependent optical selection rules. More specifically, the direct interband transitions in the vicinity of the K+ (K-) point of the hexagonal Brillouin zone are coupled to right (left) circular photon polarization states, allowing to control the properties of the material through circular polarized light.[70,71] Nonetheless the experimental works reported so far on 3R-$MoS_2$ are based on synthetic crystals and not naturally occurring minerals.[72,73] Figure 2e and 2f show a picture of 3R-molybdenite micro-crystals on the surface of a quartz rock and a optical microscopy image of the resulting mechanically exfoliated 3R-molybdenite flakes, after transferring them to a 285 nm $SiO_2$/Si surface.

Tungstenite mineral (with formula $WS_2$) is also found in nature in its 2H polytype as a (rare) micromineral. This is probably the reason why although tungstenite is naturally available, most works used synthetic $WS_2$ instead and only few examples of exfoliation of natural tungstenite have been reported in the literature. Tungstenite, similarly to molybdenite, is a n-type indirect bandgap semiconductor in bulk (~1.3 eV) that becomes a direct bandgap semiconductor when thinned down to a single-layer (with a gap value of 2 eV).[74] The spin-orbit splitting of the valence band in single-layer tungstenite reaches 430 meV, which is approximately three times larger than in single-layer molybdenite (150 meV), and thus this material is very interesting for the incipient field of spin-orbitronics.[75–78] Withers et al. isolated $WS_2$ flakes with thickness ranging from single-layer up to 4 layers by exfoliation of natural tungstenite and studied its intrinsic electronic properties by fabricating field effect devices on hexagonal boron nitride, showing a marked n-type behavior and mobilities up to 80 $cm^2V^{-1}s^{-1}$ at room temperature.[9] Figure 2g, 2h and 2i shows the crystal structure, a picture of a natural tungstenite micro-mineral on the surface of a quartz rock and an optical microscopy image of mechanically exfoliated tungstenite flakes.

Orpiment and anorpiment are two minerals with the same chemical formula ($As_2S_3$) but differing in their respective crystal lattices. Figures 2j and 2m display the crystal structure of these two materials, with orpiment belonging to the monoclinic crystal system while anorpiment is the triclinic dimorph of orpiment. [79] It is worth mentioning that bulk anorpiment mineral was identified only 10 years ago and thus there is still scarce information about this material, even in the bulk form.[80] In this bulk form, orpiment is an indirect wide bandgap semiconductor with bandgap in the range of ~2.4-2.6 eV and a intense yellow color.[79,81] Recently, Manjón et al. studied the vibrations, structure and electronic properties of orpiment under high pressure finding that pressure is able to tune its metavalent bonding thus turning orpiment from a semiconductor into an “incipient metal” with promising phase-change, thermoelectric and topological insulating properties.[82] Mortazavi and co-workers have calculated the mechanical properties and the mobility of orpiment nanosheets reporting a strong anisotropy in the mechanical properties: along one crystal direction orpiment is elastic and brittle, whereas along the perpendicular direction it shows superstretchability, similar to rubber.[83] Interestingly, Steeneken et al. isolated single- and few-layers of orpiment by mechanical exfoliation and tested their strong vibrational and mechanical anisotropy, arising from its crystal structure, finding a Young’s modulus of $E_{a\text{-}axis}$ = 79.1 ± 10.1 GPa and $E_{c\text{-}axis}$ = 47.2 ± 7.9 GPa along the two main crystalline directions.[84] This Young’s modulus anisotropy is the largest reported in the literature so far. Figures 2k and 2n show pictures of orpiment and anorpiment mineral rocks and Figures 2l and 2o are optical microscopy images of flakes exfoliated from thos minerals.

Stibnite has a similar chemical composition to orpiment and anorpiment with Sb atoms replacing the As atoms to form the structure $Sb_2S_3$. Nonetheless, the crystal structure of stibnite strongly differs from that of orpiment and anorpiment. The S atoms are coordinated in a pyramidal fashion around the Sb atoms to form a chain structure with units given by $Sb_4S_6$ ribbons (see Figure 2p). Stibnite , similarly to native selenium or tellurium, presents layers which are not fully covalently bonded inside the plane, and thus their structure resembles more to that of molecular solids like rubrene. This peculiar crystal structure has motivated works studying the in-plane anisotropy of the optical and electronic properties of stibnite and related materials, although these works relied on artificially synthesized material sources rather than in stibnite mineral.[85] In bulk stibnite is a semiconductor with a gap of ~1.6-1.7 eV (there is no consensus in the literature about the direct/indirect nature) with interest in photocatalysis and photovoltaics.[79,81,86–88] Figures 2q and 2r show a picture of a stibnite mineral rock and an optical microscopy image of flakes exfoliated trom the buk mineral.

Getchellite is a relative of the orpiment and stibnite minerals. It has a formula of $AsSbS_3$ and it has a complex crystal structure that ressembles a mixture between those of orpiment and stibnite. Very recently, Wang and co-workers reported the isolation of getchellite flakes (45 nm thick) by mechanical exfoliation of synthetic getchellite crystals and their characterization by transient absorption spectroscopy.[89] They found that $AsSbS_3$ has a direct band gap of 1.74 eV and they determined the mobility of the charge carriers of 200 $cm^2$/Vs.

P. Gehring et al. reported the exfoliation of natural kawazulite mineral (with the approximate composition $Bi_2(Te,Se)_2(Se,S)$), a naturally occurring topological insulator.[90] They isolated flakes by mechanical exfoliation with thickness down to few tens of nanometers and fabricated electronic devices with carrier mobility exceeding 1000 $cm^2V^{-1}s^{-1}$. Moreover, angle-resolved photoelectron spectroscopy show a surface state with the typical Dirac-like conical dispersion, which verifies the topological insulator behavior of natural kawazulite. Athough tin sulfide minerals like herzenbergite (with formula SnS, a puckered honeycomb structure and a band gap of ~1.0-1.1 eV) and berndtite (formula $SnS_2$, a structure similar to that of 1T $MoS_2$ polytype and a bandgap of ~2.2 eV) are found in nature, we have not found works reported on exfoliation of these natural minerals but on the exfoliation of their synthetic counterparts.[91–94]

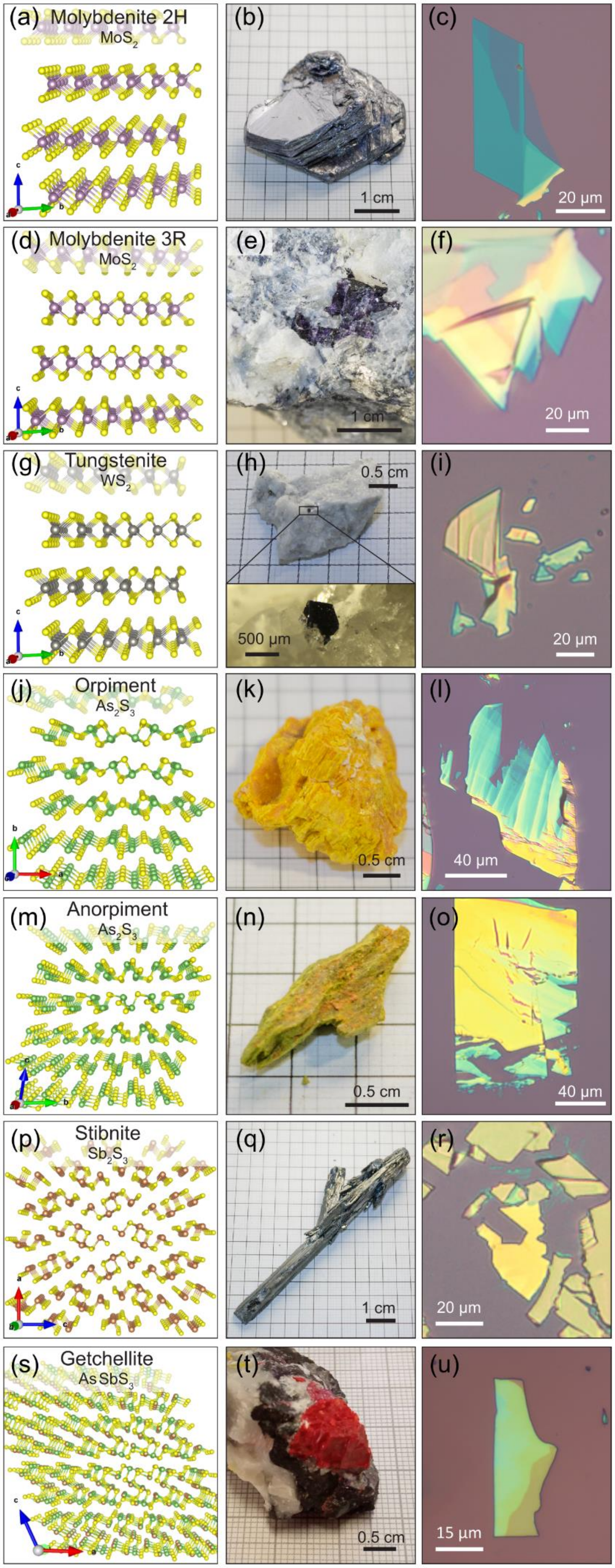

**Figure 2: Sulfide van der Waals minerals.** (a), (d), (g), (j), (m), (p) and (s) 3D representation of the crystal structure of molybdenite 2H,[155] molybdenite 3R,[155] tungstenite 2H,[156] orpiment,[157] anorpiment[80], stibnite[158] and getchellite[159,160] respectively. (b), (e), (h), (k), (n), (q) and (t) Pictures of mineral rocks of molybdenite 2H, molybdenite 3R, tungstenite 2H, orpiment, anorpiment, stibnite and getchellite, respectively. (d), (f), (i), (l), (o), (r) and (u) Optical microscopy images of few-layers flakes mechanically exfoliated from the bulk crystals shown in (b), (e), (h), (k), (n), (q) and (t) respectively.

**Sulfosalts**

The sulfosalts are a mineral family with a general formula $A_mB_nS_p$ with A being commonly copper, lead, silver or iron, B being semimetals like arsenic, antimony or bismuth or metals like tin and S being sulfur. Among the layered sulfosalts teallite is probably the simplest one. It has a formula of $PbSnS_2$ and its crystal structure largely resembles that of the compound SnS (a puckered honeycomb lattice very similar to that of black phosphorus but with different atomic species in the unit cell). Figure 3a shows a 3D representation of the crystal structure of teallite. This particular crystal structure might result interesting for applications looking for semiconducting materials with strongly anisotropic in-plane properties. In bulk teallite is a semiconductor with ~1.6 eV of direct band gap.[95] Recently, teallite has been mechanically exfoliated and its Raman spectra has been measured as a function of the incident light polarization at different temperatures fingidn a strong linear-dichroism arising from its puckered structure.[96] Shu et al. reported the growth of ultrathin synthetic tellite, as thin as 2.4 nm, by chemical vapor deposition finding a large anisotropy ratio in the charge carrier mobility (1.8) and photoresponse (1.25) and photoresponsivity up to 20 A/W and response times of milliseconds.[97] Also, shear force based liquid phase exfoliation of teallite has been demonstrated achieving suspensions with 11-22 layers in thickness and the exfoliated material has been subsequently employed in electrocatalytic reactions.[98] Figure 3b and 3c shows a picture of a teallite mineral rock and an optical microscopy image of a teallite flake mechanically exfoliated from the bullk mineral.

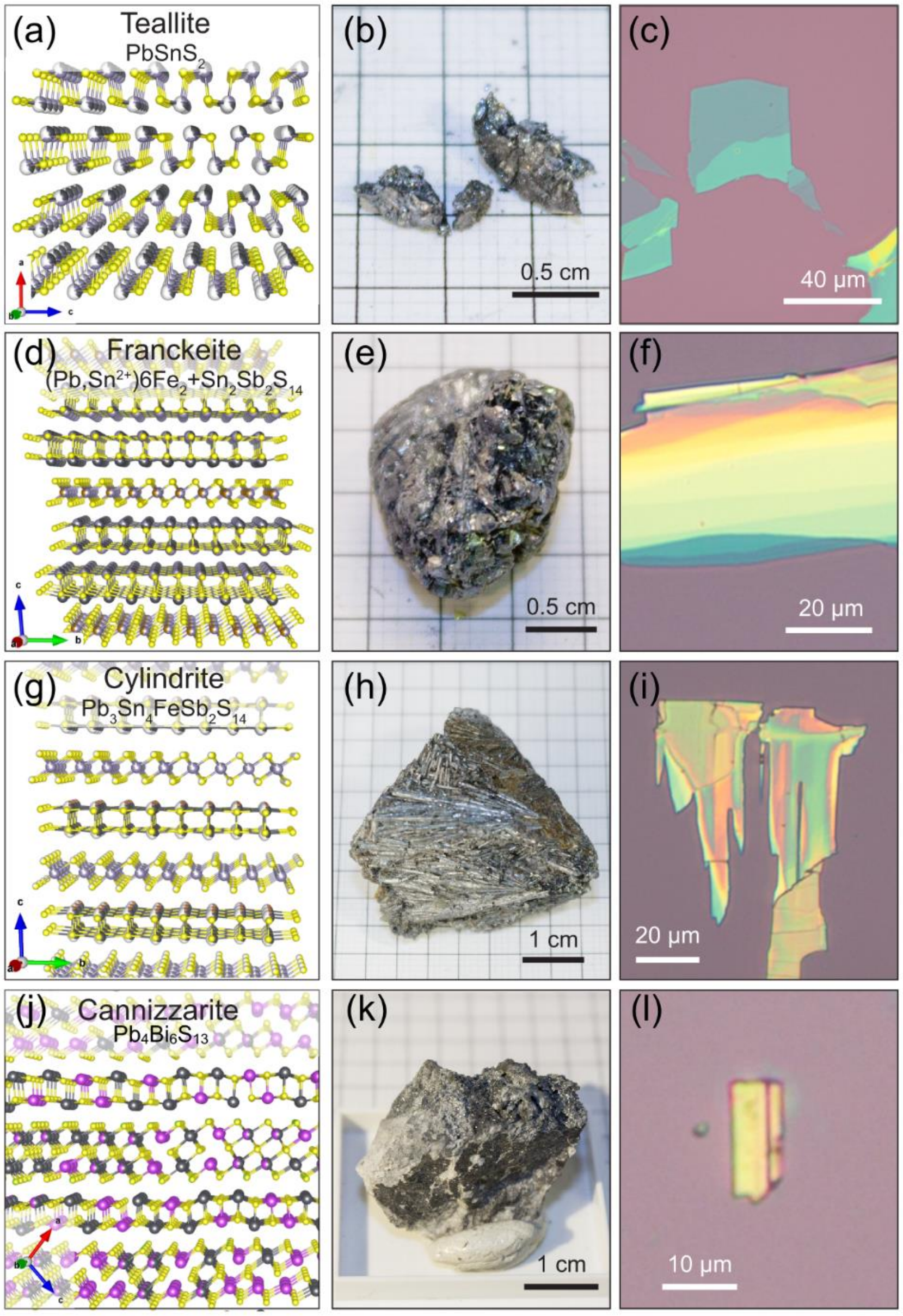

**Figure 3: Sulfosalt van der Waals minerals.** (a), (d), (g) and (j) 3D representation of the crystal structure of teallite,[161] frackeite,[162] cylindrite[163] and cannizzarite[164] respectively. (b), (e), (h) and (k) Pictures of mineral rocks of teallite, frackeite, cylindrite and cannizzarite, respectively. (d), (f), (i) and (l) Optical microscopy images of few-layers flakes mechanically exfoliated from the bulk crystals shown in (b), (e), (h) and (k) respectively.

## Phase segregated sulfosalts: naturally occurring van der Waals superlattices

Some members of the sulfosalt family are composed of alternating layers with dissimilar chemical composition (i.e. they constitute naturally occurring van der Waals superlattices [99]). Some illustrative examples of this sub-family are the franckeite, the cylindrite and the cannizzarite.

Franckeite bulk mineral (with an approximate formula $Pb_5Sn_3Sb_2S_{14}$) has been recently exfoliated down to a single unit cell.[100–104] The crystal structure is composed of segregated Sn-rich, pseudo-hexagonal layers (with $SnS_2$ like structure) alternated to Pb-rich, pseudo-tetragonal layers (with PbS like structure) with a van der Waals gap between them (see Figure 3d). Molina-Mendoza et al, Velický et al. and Ray et al. demonstrated field-effect devices, photodetectors and solar-cells with mechanically exfoliated flakes.[100,101,103] These first works demonstrated that this exfoliated material is a p-type semiconductor with a bandgap value of ~0.6 eV [100] similarly to what has been reported for bulk franckeite.[81] Liquid phase exfoliation with different solvents have been used to produce colloidal suspensions of flakes with thickness down to 4 unit cells.[98,100,101,105] Field-effect devices with performances very similar to that of mechanically exfoliated based devices, that show a strong p-type doping and a field-effect modulation up to a factor of 10, have been demonstrated from the liquid phase exfoliation franckeite suspensions through dielectrophoresis based assembly.[105] More recent works characterized the optical properties of exfoliated frankeite nanosheets[104,106] and it has ben shown how covalent thiol-ene-like "click" chemistry can be used to decorate franckeite.[107] Also, Frisenda and co-workers have recently studied the spontaneous symmetry breakdown in frankeite resulting from a spatial modulation of the van der Waals interaction between layers due to the $SnS_2$-like and PbS-like lattices incommensurability. In fact, although franckeite superlattice is composed of a sequence of isotropic 2D layers, it exhibits a spontaneous rippling that makes the material structurally anisotropic leading to an inhomogeneous in-plane strain profile and anisotropic electrical, vibrational, and optical properties.[108] Figures 3e and 3f show a picture of a franckeite mineral rock and an optical microscopy image of a frankeite flake mechanically exfoliated from the bulk crystal.

Cylindrite (with an approximate formula $Pb_3Sn_4FeSb_2S_{14}$) is a mineral closely related to franckeite. It received its name because it often occurs as cylindrical crystals made up of rolled sheets. The structure, similar to franckeite, presents alternating segregated Sn-rich and Pb-rich layers (see Figure 3g). In bulk cylindrite is a semiconductor with a band gap of 0.65 eV, very similar to that of franckeite.[81] Interestingly, bulk cylindrite present intrinsic magnetic interactions, compatible with a spin glass-like system, which are assumed to be originated by its iron content.[109,110] Recently, Niu et al. reported the isolation of thin flakes of cylindrite (down to 9-10 layers) by mechanical and liquid phase exfoliation, finding that the flakes of this material are heavily doped p-type semiconductors with a narrow gap (<0.85 eV). They also found intrinsic magnetic interactions (the magnetization shows an anomaly at ~5K that can be attributed to a slowing down of the spin dynamics in magnetically disordered systems, the hallmark of spin glass-like behavior) that are preserved even in the exfoliated nanosheets. Figures 3h and 3i show a picture of a cylindrite mineral rock and an optical microscopy image of cylindrite flakes exfoliated from the bulk rock.

Cannizzarite (with an approximated formula $Pb_4Bi_6S_{13}$) is a rare micro-mineral formed by alternating stacks of tetragonal PbS-like and hexagonal layers $Bi_2S_3$ like (see the 3D representation of the crystal structure in Figure 3j).[109,110] Very little has been reported so far on this mineral, even in its bulk form. Figure 3k is a picture of a quartz rock covered by small micro-crystallites of cannizzarite, these crystals can be lifted up from the surface using a viscoelastic Gel-Film (by Gel-Pak) stamp. Figure 3l shows a cannizzarite flake after picking it up from the bulk rock and transfer it to a $SiO_2$/Si substrate by an all-dry deterministic transfer method.

**Oxides**

The oxides are minerals in which the oxide anion ($O^{2-}$) is bonded to one or more metal ions. Complex oxides such as silicates, carbonates and phosphates are traditionally classified separately. Figures 4a and 4b shows an example of one naturally occurring oxide: valentinite (with formula $Sb_2O_3$ and orthorhombic structure) occurs as a weathering product of stibnite and other antimony-based minerals. It is a semiconductor with a wide band gap of ~3.3 eV (no consensus about the direct/indirect nature of the gap) and dielectric constant of ~3 that has recently been used in energy storage applications.[111–114] Other example of oxide mineral is the birnessite (see Figures 4c and 4d), with approximate formula $(Na,Ca,K)_{0.6}(Mn^{4+},Mn^{3+})_2O_4 \cdot 1.5H_2O$, that has been exfoliated by liquid phase expoliation technique.[13,115] In bulk birnessite is a wide band gap semiconductor with an indirect gap at ~2.1 eV and a direct gap at ~2.7 eV and a dielectric constant of ~5[116–118] and interesting for supercapacitors and photoelectrochemistry applications.[117–122] Moreover, there are theoretical predictions that exfoliated monolayer birnessite should exhibit intrinsic ferromagnetism with a Curie temperature of 140 K.[111] We are not aware of other experimental works reporting the exfoliation of other naturally occurring oxide minerals.

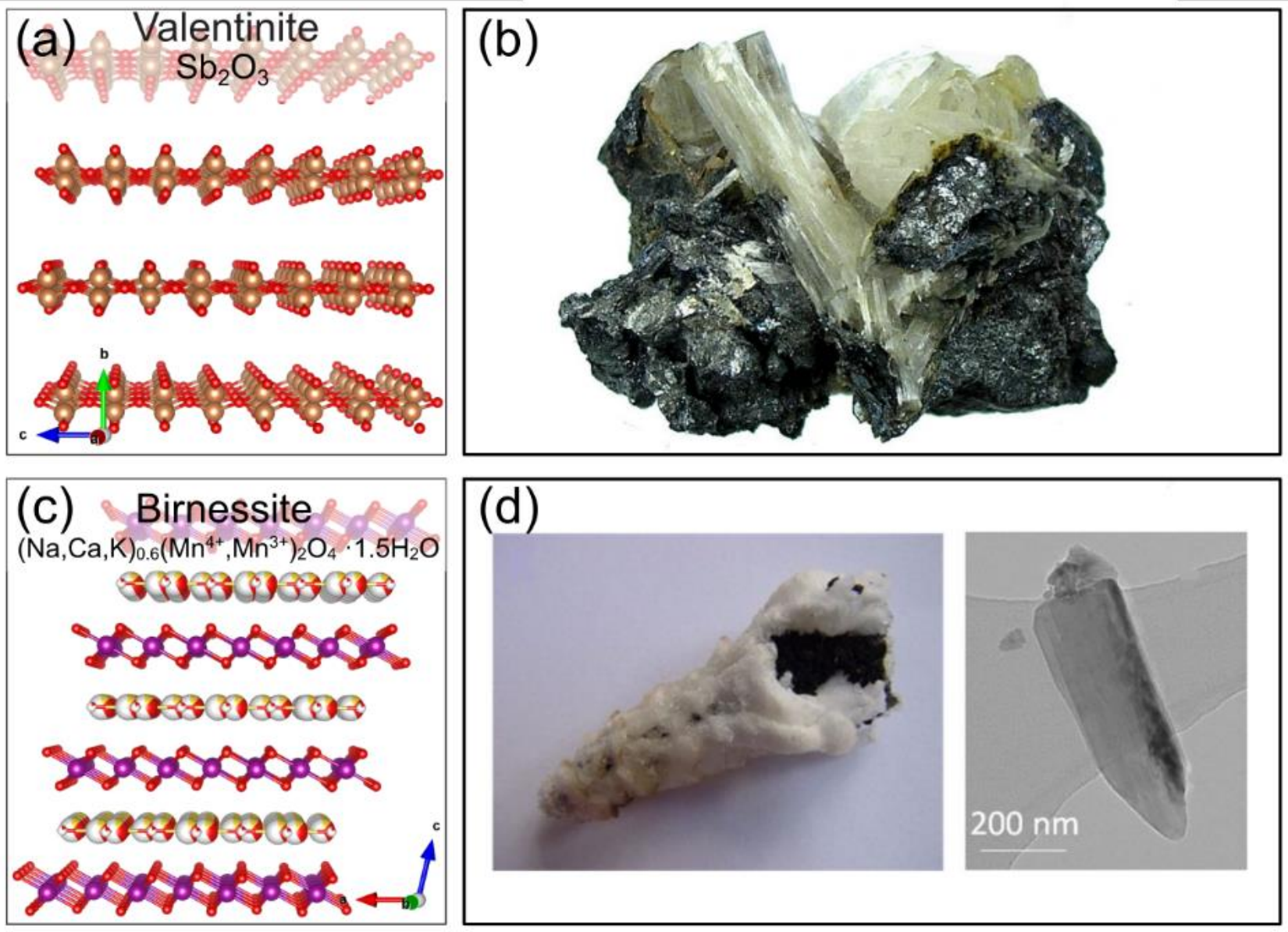

**Figure 4: Oxide van der Waals minerals.** (a) and (c) 3D representation of the crystal structure of valentinite[165] and birnessite[166] respectively. (b) and (d) Pictures of mineral rocks of valentinite[167] and birnessite,[13] respectively. (d) includes a transmission electron microscopy image of an exfoliated bernissite flake. Panel (d) reproduced from Ref.[13] with permission.

### Silicates

Silicates is a family of largely abundant minerals made out of silicate groups. These materials are wide band gap insulators and some of them are typically used as dielectrics for capacitors in the semiconductor industry.

#### Nesosilicates

Nesosilicates are silicates that have $SiO_4$ tetrahedra that are isolated and connected by interstitial cations. Kyanite is an illustrative example of nesosilicate mineral with formula $Al_2SiO_5$ and intense blue color. Its structure is composed of "staircases" of Al octahedra linked by Si tetrahedra (see Figure 5a).[123] It is a wide gap insulator with a direct band gap of ~5-6 eV and a dielectric constant of ~9.[124–126] High resolution frictional AFM images of the surface of bulk kyanite has been recently reported. These friction measurements performed along the [001] and [010] directions on the kyanite (100) face provide similar friction coefficients $\mu \approx 0.10$.[123] Figure 5b shows a picture of a kyanite mineral rock and Figure 5c is an optical microscopy image of a kyanite flake exfoliated from the bulk mineral.

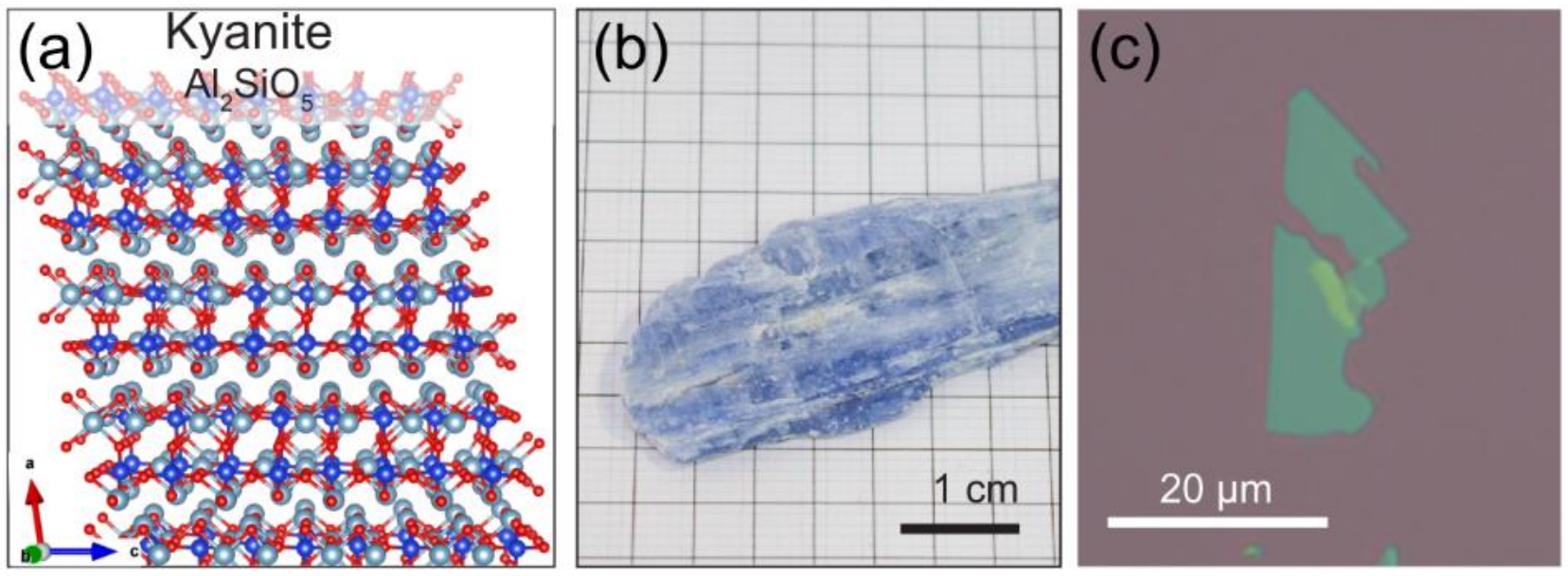

**Figure 5: Nesosilicate van der Waals minerals.** (a) 3D representation of the crystal structure of kyanite.[168] (b) Picture of a mineral rock of kyanite. (c) Optical microscopy image of a flake of kyanite mechanically exfoliated from the bulk crystal shown in (b).

## Phyllosilicates: clays, micas and chlorites

These materials are silicates with a 2:5 Si to O ratio. This large family of silicates includes the clays, micas and chlorites and all the members of this family are hydrated with either water or hydroxyl groups attached.

Clays are hydrous aluminum or magnesium phyllosilicates. From clays one can obtain organoclays, in which the original interlayer cations (such as $Na^+$) are exchanged with organocations (for example quaternary alkylammonium ions) leading to a different layer separation. Previous works showed that clay minerals can be exfoliated in liquid phase with the help of large polymers that intercalate in the crystal structure separating the layers. Talc (with formula $Mg_3Si_4O_{10}(OH)_2$) is an illustrative example of the clays family. It is a wide band gap insulator (direct gap of 5.2 eV) with high dielectric constant (9.4).[127,128] The isolation of talc nanosheets with thickness down to one single-layer has been recently reported by mechanical exfoliation [127] and liquid phase exfoliation has been also used to prepare suspensions of talc nanosheets with an average of 9 layers.[129] Mechanically exfoliated talc flakes have been also used as substrates or encapsulating layers to fabricate graphene-based electronic devices where the electronic interaction between the talc and the graphene are exploited (the dielectric constant of talc is higher than other encapsulation layers like hexagonal boron nitride) to modify the performance of the devices.[130,131] Figures 6a, 6b and 6c show the representation of the crystal structure of talc, a picture of a mineral rock and an optical microscopy image of a exfoliated talc flake, respectively.

Apart from talc, liquid phase exfoliation has been demonstrated to be a powerful route to isolate suspensions of other members of the clay family.[129] Figures 6d and 6e shows the crystal structure and a picture of bulk vermiculite (general formula $Mg_{0,7}(Mg,Fe,Al)_6(Si,Al)_8O_{20}(OH)_4·8H_2O$, an insulator with a dielectric constant of ~5)[132] and expanded vermiculite that can be readily exfoliated by liquid phase exfoliation technique.[13]

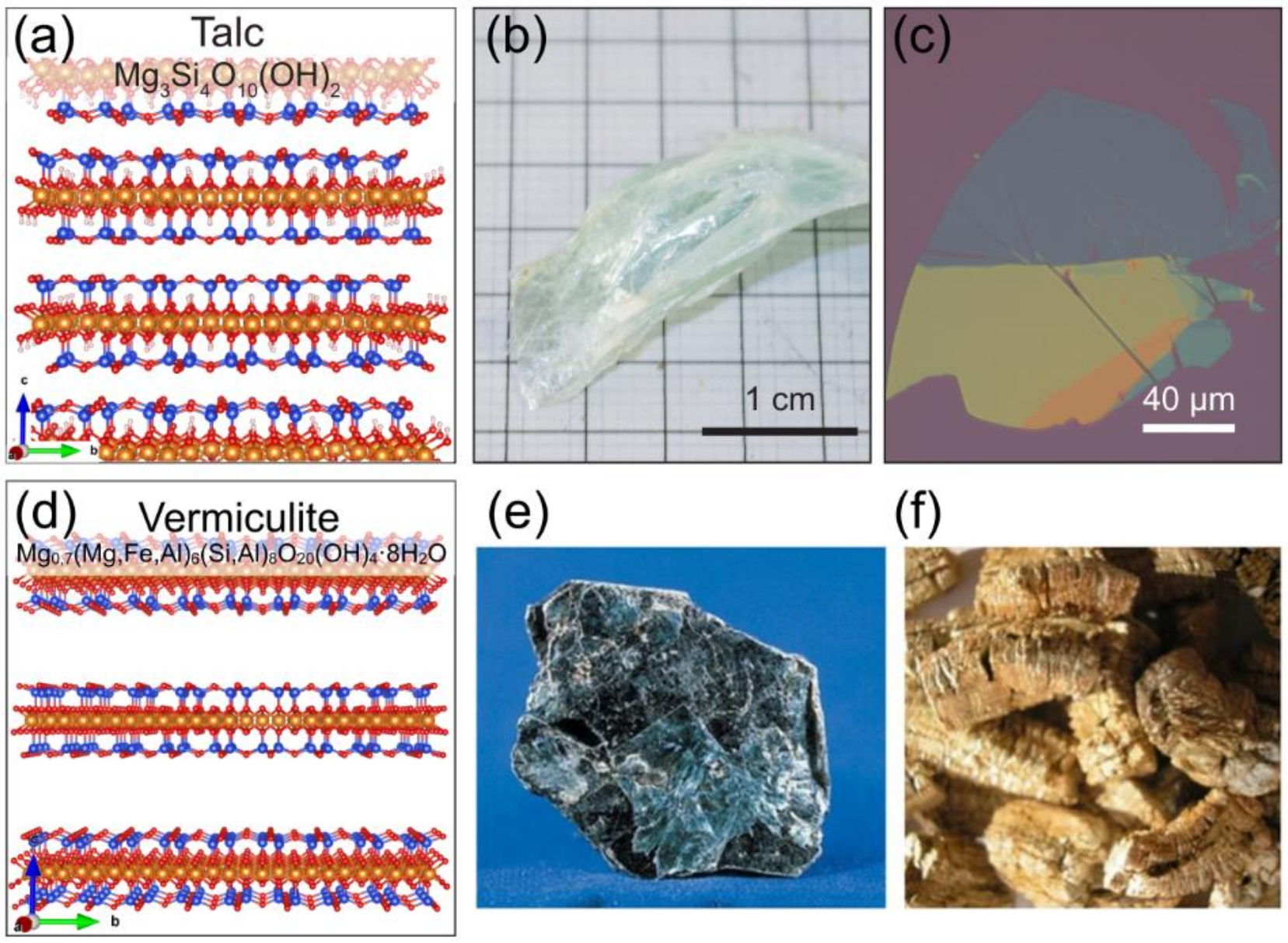

**Figure 6: Phyllosilicate clay van der Waals minerals.** (a) and (d) 3D representation of the crystal structure of talc [169] and vermiculite.[170] (b) and (e) Pictures of mineral rocks of talc and vermiculite,[13] respectively. (c) Optical microscopy image of a talc flake, mechanically exfoliated from the bulk crystal shown in (b). (f) Picture of thermally expanded vermiculite.[13] Panels (e) and (f) reproduced from Ref.[13] with permission.

The mica family is formed by very closely related layered silicates such as muscovite (with formula $KAl_2(AlSi_3)O_{10}(OH)_2$), biotite (with formula $K(Mg,Fe)_3(AlSi_3)O_{10}(OH)_2$), lepidolite (with formula $K(Li,Al)_{2-3}(AlSi_3)O_{10}(OH)_2$), phlogopite (with formula $KMg_3(AlSi_3)O_{10}(OH)_2$). Most of the previous works on exfoliated micas are focused on muscovite mica, a wide band insulator (5.1 eV of direct gap) with a large dielectric constant (~10) in its bulk form.[133,134] The exfoliation of layered materials with stronger interlayer forces has rarely been reported and mica is one of these few cases. In fact, muscovite mica has been successfully exfoliated down to a single-layer by mechanical exfoliation and their optical properties [10,12] as well as their mechanical properties have been studied finding an optical method to determine its thickness and a Young's modulus of ~200 GPa, higher than that of bulk muscovite (~175 GPa).[11] Monolayer muscovite mica nanosheets can also be obtained by liquid phase exfoliation as well as by weakening its layer attractions (enlarging the basal spacing by intercalation) prior to sonication.[135] Mechanically exfoliated muscovite mica flakes have been used as dielectric to fabricate graphene transistors with improved performance due to the high dielectric constant of mica and the atomically flat surface achieved after mechanical exfoliation.[136] Interestingly, the hydrophilic nature of muscovite mica in combination with the impermeability of graphene has opened up the possibility of studying the structure and the electronic properties of water confined a the mica/graphene interface.[137–139]

Figure 7 summarizes the crystal structure sketch, optical images of the bulk minerals and microscopy images of the corresponding exfoliated flakes for muscovite, biotite, lepidolite and fluogopyte micas.

Members of the chlorite family have a great range in composition resulting in a large variety of physical, optical, and X-ray properties. Clinochlore is an illustrative example of the chlorite family. This mineral, with formula $(Mg,Fe^{2+})_5Al(Si_3Al)O_{10}(OH)_8$, resembles the layered minerals of the mica family but it presents a characteristic greenish color and a more plastic response to bending. To our knowledge, there are no reports of exfoliated chlorite materials in the literature so far.  Figure 8 summarizes the crystal structure sketch, the picture of a clinochlore rock and an ptical microscopy image of a mechanically exfoliated clinochlore flake.

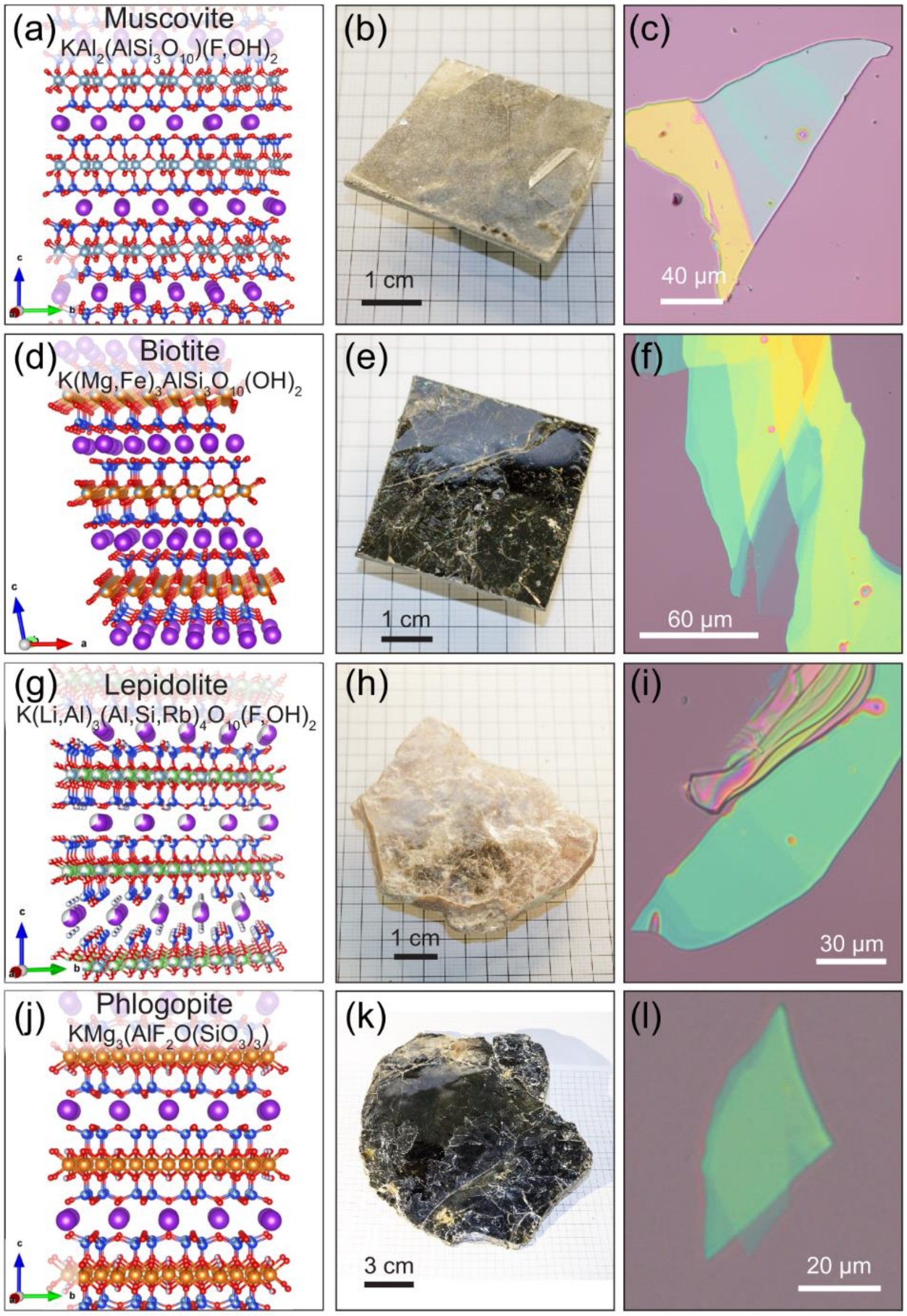

**Figure 7: Phyllosilicate mica van der Waals minerals.** (a), (d), (g) and (j) 3D representation of the crystal structure of muscovite,[171] biotite,[172] lepidolite[173] and phlogopite[174] respectively. (b), (e), (h) and (k) Pictures of mineral rocks of muscovite, biotite, lepidolite and phlogopite, respectively. (d), (f), (i) and (l) Optical microscopy images of few-layers flakes mechanically exfoliated from the bulk crystals shown in (b), (e), (h) and (k) respectively.

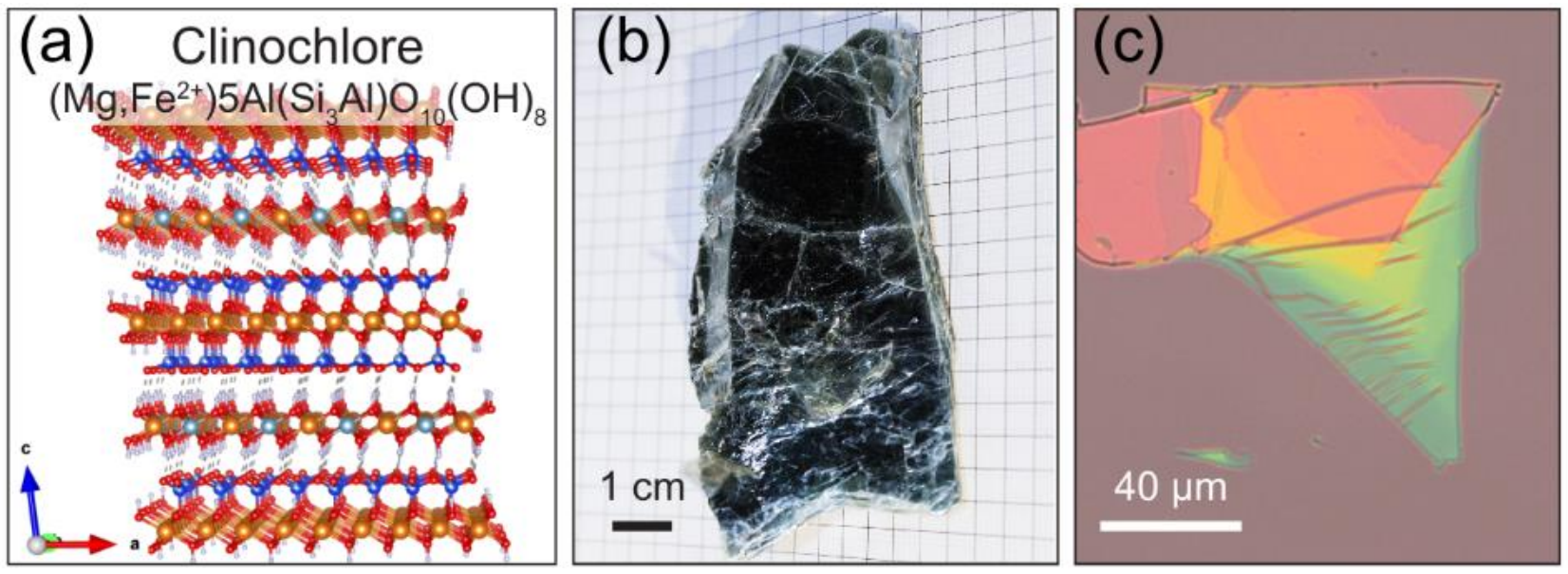

**Figure 8: Phyllosilicate chlorite van der Waals minerals.** (a) 3D representation of the crystal structure of clinochlore.[175] (b) Picture of a mineral rock of clinochlore. (c) Optical microscopy image of a clinochlore flake mechanically exfoliated from the bulk crystal shown in (b).

## Phosphates

Phosphates are minerals that contain the tetrahedrally coordinated phosphate ($PO_4^{3-}$) anion. Vivianite is a hydrated iron phosphate with an approximate formula $Fe^{2+}Fe_2^{2+}(PO_4)_2 \cdot 8H_2O$. Vivianite is known to be sensitive to visible light exposure which leads to a marked change in its color from colorless/pale green to dark green/brown. Bulk vivianite has been used as natural electron donor to effectively dechlorinate a variety of chlorinated organics, the principal and most frequently found contaminants in soil and groundwater that generates significant environmental problems.[132] We also are not aware of any works in the literature reporting the exfoliation of vivianite in atomically thin flakes. Moreover, we could not find experimental reports on the band gap of vivianite but a recent theoretical calculation suggest that vivianite would present an indirect band gap in the range of ~3.3-4.6 eV and a paramagnetic ground state.[140,141] Figure 9 summarizes the crystal structure sketch, the picture of the vivianite rock and an optical microscopy image of a mechanically exfoliated vivianite flake.

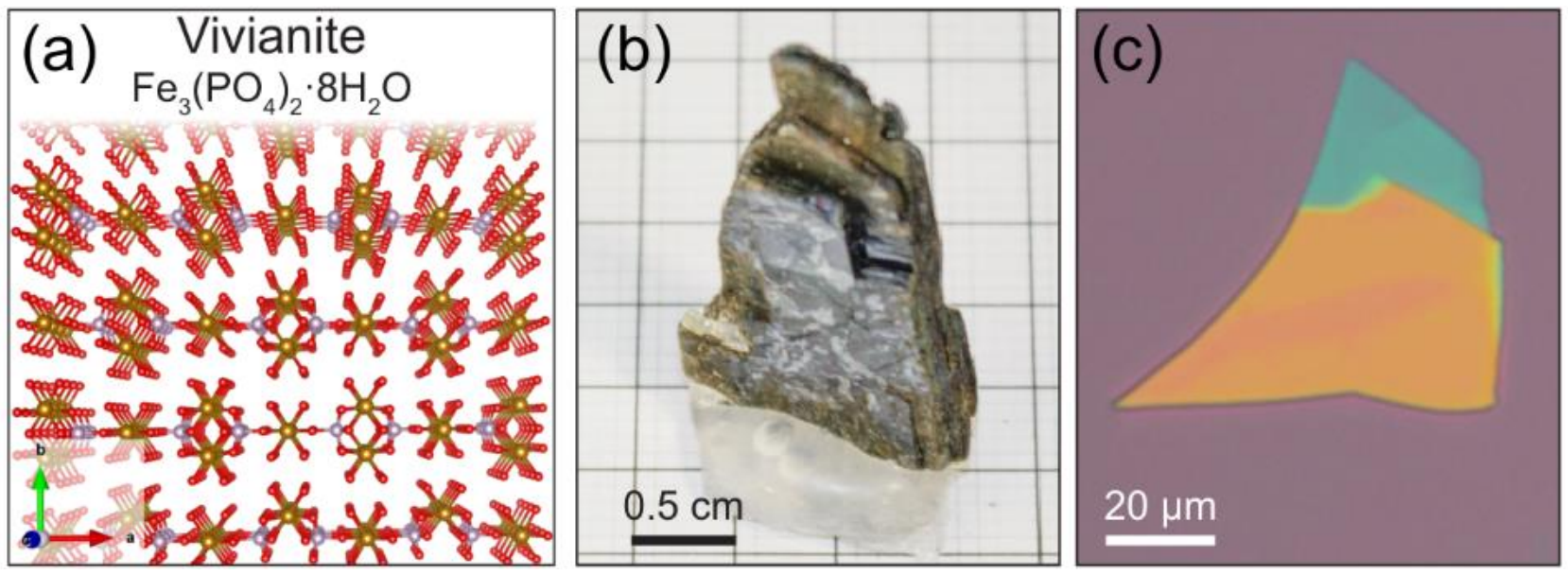

**Figure 9: Phosphate van der Waals minerals.** (a) 3D representation of the crystal structure of vivianite.[176] (b) Picture of a mineral rock of vivianite. (c) Optical microscopy image of a vivianite flake mechanically exfoliated from the bulk crystal shown in (b).

## Carbonates

Carbonates are the minerals containing the carbonate ion ($CO_3^{2-}$). Malachite is a copper carbonate hydroxide with a formula $Cu_2CO_3(OH)_2$. It has a dielectric constant on ~7 but we could not find information about its band structure.[126] Figure 10 summarizes the crystal structure sketch, the picture of the malachite rock and an optical microscopy image of a mechanically exfoliated malachite flake. Unfortunately, little can be found in the literature, apart from its crystal structure,[142] even for the properties of bulk malachite.

Note that although this overview of natural van der Waals minerals is most likely far from being complete, as we expect that there are many other layered mineral families not discussed in this perspective, we believe that it will constitute a good starting point to motivate the scientific community working on 2D materials to study these natural materials. Table 1 summarizes the main information highlighted over this perspective for the different minerals discussed here.

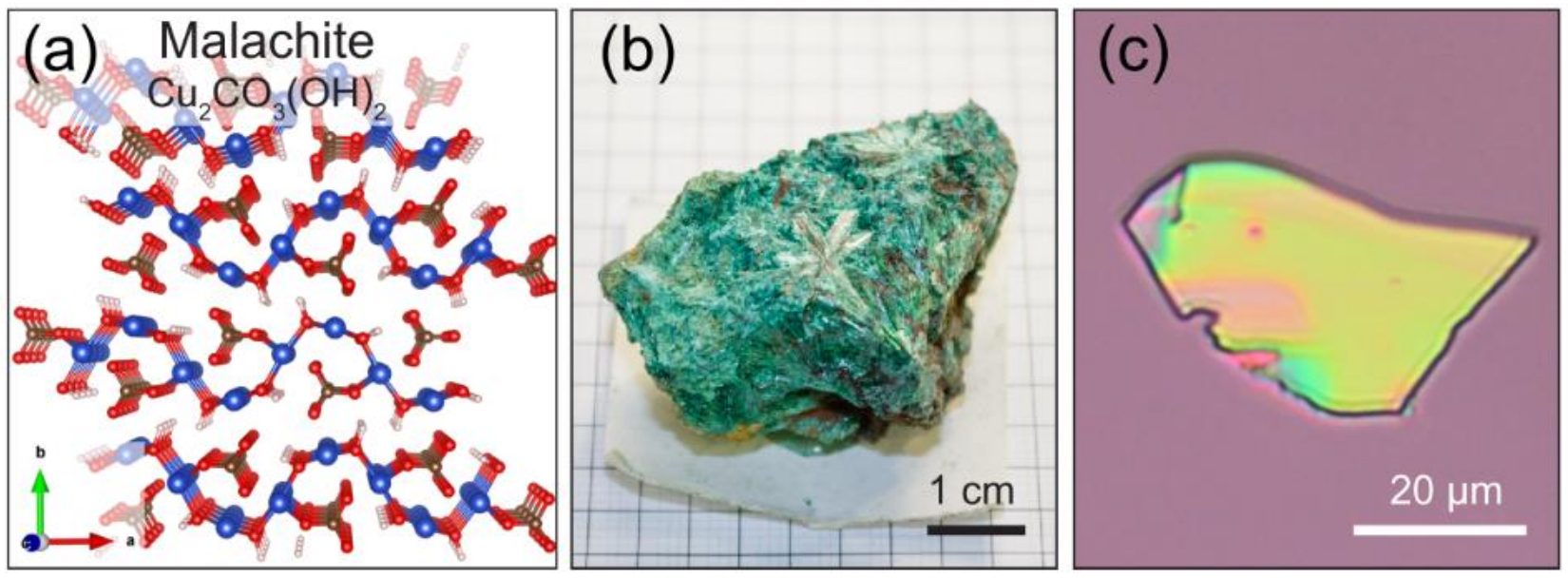

**Figure 10: Carbonate van der Waals minerals.** (a) 3D representation of the crystal structure of malachite.[142] (b) Picture of a mineral rock of malachite. (c) Optical microscopy image of a flake of malachite mechanically exfoliated from the bulk crystal shown in (b).

| Group | | | Mineral | Formula | Electronical behavior | Band gap (eV) |
|---|---|---|---|---|---|---|
| Elemental | | | Graphite | C | Semimetal | 0 |
| | | | Bismuth | Bi | Metal | 0 |
| | | | Antimony | Sb | Semimetal | 0 |
| | | | Selenium | Se | Semiconductor | 2 (d) |
| | | | Tellurium | Te | Semiconductor | 0.31 (i) |
| Sulfides | | | 2H- Molybdenite | $MoS_2$ | Semiconductor | 1.3 (i) |
| | | | 3R- Molybdenite | $MoS_2$ | Semiconductor | 1.3 (i) |
| | | | 2H- Tungstenite | $WS_2$ | Semiconductor | 1.3 (i) |
| | | | Orpiment | $As_2S_3$ | Semiconductor | ~2.4-2.6 (i) |
| | | | Anorpiment | $As_2S_3$ | Semiconductor | N/A |
| | | | Stibnite | $Sb_2S_3$ | Semiconductor | ~1.6-1.7 (?) |
| | | | Getchellite | $AsSbS_3$ | Semiconductor | 1.74 (d) |
| Sulfosalts | | | Teallite | $PbSnS_2$ | Semiconductor | ~1.6 (d) |
| | | | Franckeite | $Pb_5Sn_3Sb_2S_{14}$ | Semiconductor | ~0.6 (?) |
| | | | Cylindrite | $Pb_3Sn_4FeSb_2S_{14}$ | Semiconductor | 0.65 (?) |
| | | | Cannizzarite | $Pb_4Bi_6S_{13}$ | N/A | N/A |
| Oxides | | | Valentinite | $Sb_2O_3$ | Insulator | ~3.3 (?) |
| | | | Birnessite | $(Na,Ca,K)_{0.6}(Mn^{4+},Mn^{3+})_2O_4 \cdot 1.5H_2O$ | Insulator | ~2.1 (i) and ~2.7 (d) |
| Silicates | Nesosilicates | | Kyanite | $Al_2SiO_5$ | Insulator | ~5-6 (d) |
| | Phyllosilicates | Mica | Muscovite | $KAl_2(AlSi_3)O_{10}(OH)_2$ | Insulator | 5.1 (d) |
| | | | Biotite | $K(Mg,Fe)_3(AlSi_3)O_{10}(OH)_2$ | Insulator | N/A |
| | | | Lepidolite | $K(Li,Al)_{2-3}(AlSi_3)O_{10}(OH)_2$ | Insulator | N/A |
| | | | Phlogopite | $KMg_3(AlSi_3)O_{10}(OH)_2$ | Insulator | N/A |
| | | Chlorite | Clinochlore | $(Mg,Fe^{2+})_5Al(Si_3Al)O_{10}(OH)_8$ | Insulator | N/A |
| | | Clay | Talc | $Mg_3Si_4O_{10}(OH)_2$ | Insulator | 5.2 (d) |
| | | | Vermiculite | $Mg_{0,7}(Mg,Fe,Al)_6(Si,Al)_8O_{20}(OH)_4 \cdot 8H_2O$ | Insulator | N/A |
| Carbonates | | | Malachite | $Cu_2CO_3(OH)_2$ | N/A | N/A |
| Phosphates | | | Vivianite | $Fe^{2+}Fe_2^{2+}(PO_4)_2 \cdot 8H_2O$ | Insulator | ~3.3-4.6 (i) |

Table 1: Summary of the different naturally occurring van der Waals materials discussed in this perspective.

## Conclusions and discussion

In summary, we provided an overview over different mineral families containing members with layered structure that can be exfoliated by mechanical exfoliation. We discussed about the basic electronic and structural characteristics of these materials and we illustrate how thin flakes can be prepared by mechanical exfoliation of the bulk minerals. Most of the minerals discussed in this Perspective are easily available in specialized mineral shops or online auction sites at reasonable price (typically <50$ per mineral) making it easier the access to exotic van der Waals materials whose synthetic counterparts might be much more expensive (typically ~500$ per crystal) or even not available for purchase (e.g. some sulfosalts). The study of naturally occurring layered materials, however, can present some challenges. One of the most important ones is the purity/quality of the natural van der Waals mineral as compared to the syntheric counterpart. One might naively think that natural minerals are more prone to have impurities than synthetic ones. However, in the literature we can find two clear counterexamples: plenty of high-quality results for graphene and $MoS_2$ are obtained with mechanically exfoliated graphite and molybdenite. In the case of $MoS_2$, for example, the highest mobility reported to date has been obtained with $MoS_2$ flakes extracted from natural molybdenite (SPI source).[6] We believe, however, that minerals with a complex structure and chemical composition (e.g. the sulfosalts) will suffer more from un-controlled impurities than their synthetic counterparts. Another important issue will be to systematically study the optical and electrical properties of mineral coming from different mines to assess the impact of the parenting mineral source on the properties of the exfoliated 2D materials. We would like to note that this is a general important issue, not specific of naturally occurring van der Waals materials but of 2D materials research as a whole. Indeed one would expect that the source of the parenting bulk layered material (even synthetic ones) might have a strong impact in the properties of the exfoliated flakes produced from it. Systematic studies trying to correlate the properties of exfoliated materials produced from different sources are still scarce.[143–145] Many of these layered minerals are almost unexplored so far and we believe that this overview can constitute a necessary first step to trigger further works on exfoliation of naturally occurring layered minerals to produce 2D materials.

## Methods

*Materials:* the minerals used in this work come from the private collection of A.C-G. Unfortunately we cannot track back the original supplier for all the materials as these pieces have been gathered along more than 10 years, in mineral shops around the world and online mineral auctions (e.g. eBay or e-Rocks).

*Exfoliation of the minerals:* mineral bulks were mechanically exfoliated with Nitto tape (Nitto SPV 224) and then transferred onto Gel-Film (Gel-Pak, WF 4× 6.0 mil), which is a commercially available polydimethylsiloxane (PDMS) substrate. The resulting flakes on the surface of the Gel-Film were transferred onto a $SiO_2$/Si substrate (with 285 nm of $SiO_2$ capping layer) by means of an all-dry deterministic placement metod.[37]

*Optical microscopy imaging of exfoliated flakes:* Optical microscopy images have been acquired with two upright metallurgical microscopes a Motic BA MET310-T and a Nikon Eclipse CI equipped with an AM Scope mu1803 and a Canon EOS 1200D camara, respectively.

*3D representation of the crystal structures:* The 3D representations on the crystal structures included in the figures along the manuscript were produced with VESTA software [146] using the crystallographic data from the cited references in the figure captions. Note that most of the crystallographic data of those references can be directly found in form of .CIF files in the American Mineralogist Crystal Structure Database, making straightforward its 3D representation[146].

**Author Contribution**

AC-G designed and supervised the work and drafted the first version of the manuscript. RF, YN, PG and MM exfoliated the minerals and characterized the resulting flakes and contributed to the elaboration of the last version of the manuscript.

**ACKNOWLEDGEMENTS**

We thank Nikos Papadopoulos for interesting discussions about minerals and for pointing out the existence of getchellite to us. We also would like to thank the staff of IGE Minerales shop and Museo Geominero of Madrid. This project has received funding from the European Research Council (ERC) under the European Union's Horizon 2020 research and innovation programme (grant agreement n° 755655, ERC-StG 2017 project 2D-TOPSENSE) and the European Union's Horizon 2020 research and innovation programme under the Graphene Flagship (grant agreement number 785219, GrapheneCore2 project and grant agreement number 881603, GrapheneCore3 project). R.F. acknowledges the support from the Spanish Ministry of Economy, Industry and Competitiveness through a Juan de la Cierva-formación fellowship 2017 FJCI-2017-32919 and the grant MAT2017-87072-C4-4-P.